\documentclass[12pt]{iopart}
\usepackage{epsfig}
\newcommand{\de}{\partial}
\newcommand{\be}{\begin{equation}}
\newcommand{\ba}{\begin{eqnarray}}
\newcommand{\ea}{\end{eqnarray}}
\newcommand{\ee}{\end{equation}}

\newcommand{\f}{\frac}
\newcommand{\s}{\sqrt}
\newcommand{\vp}{\varphi}

\newcommand{\ap}{\alpha}

\newcommand{\ddd}{\cdot\cdot\cdot}

\newcommand{\la}{\langle}
\newcommand{\lb}{\rangle}

 \def\de{\partial}

 \def\f {\frac}
 
 \def\ap{\alpha}

 \def\ddd{\cdot\cdot\cdot}

 \def\la{\langle}
 \def\lb{\rangle}

\begin{document}

\topical{Entanglement entropy from a holographic viewpoint}

\author{Tadashi Takayanagi}

\address{\vspace{5mm}
Yukawa Institute for Theoretical Physics,
Kyoto University, Kyoto 606-8502, Japan\\ \vspace{5mm}
Kavli Institute for the Physics and Mathematics of the Universe,
University of Tokyo, Kashiwa, Chiba 277-8582, Japan\\ \vspace{5mm}
}
\ead{takayana@yukawa.kyoto-u.ac.jp}
\begin{abstract}
The entanglement entropy has been historically studied by many authors
in order to obtain quantum mechanical interpretations of the
gravitational entropy. The discovery of AdS/CFT
correspondence leads to the idea of holographic entanglement entropy, which is a
clear solution to this important problem in gravity. In this article, we would like to give a
quick survey of recent progresses on the holographic entanglement entropy. We focus on its  gravitational aspects, so that it is comprehensible to those who are familiar with general relativity and basics of quantum field theory.

\end{abstract}

\maketitle

\section{Introduction}

One of the most mysterious and fascinating aspects in general relativity is the existence of
black holes. They are peculiar only to theories with gravity. Among all important properties, a black hole has its own entropy, given by the Bekenstein-Hawking formula
\be
S_{BH}=\frac{\mbox{Area}(\Sigma)}{4G_N}, \label{BHF}
\ee
where $\Sigma$ is the horizon and $G_N$ is the Newton constant \cite{BeHa}. This formula (\ref{BHF}) suggests us that the degrees of freedom contained in a certain region in gravity is actually proportional to its area instead of the volume. This observation developed into the idea of holography (or holographic duality) \cite{Hol}. The holographic principle argues that a gravitational
theory in a $d+2$ dimensional spacetime $M$ is equivalent to a non-gravitational theory on a $d+1$ dimensional spacetime $\de M$, which is the boundary of $M$. The latter theory is
typically described by a quantum many-body system. A concrete example of holography was later obtained in string theory and this is called the AdS/CFT correspondence
\cite{Malda,GKPW} (for a review see \cite{adsreview}). This is
the particular case of holography where the gravity lives in a spacetime with a negative cosmological constant.

The original Bekenstein-Hawking formula (\ref{BHF}) relates the area of horizon,
which is a geometric data, to the entropy, which is a quantum mechanical quantity. This correspondence between a geometric quantity and a microscopic data is the key concept of holography. Therefore it is natural to expect such relations for more general observables.
In particular, we can ask what is the holographic dual of the areas of more general surfaces in a gravitational theory. There has been progresses in this direction recently by employing the idea of holography. The upshot is that the area of a minimal surface in a (Euclidean) gravitational theory corresponds to the entanglement entropy in its dual non-gravitational theory \cite{RuTa,RuTaL}. This is simply summarized into the formula
\be
S_A=\f{\mbox{Area}(\gamma_A)}{4G_N}, \label{HEE}
\ee
where $S_A$ is the entanglement entropy for the subsystem $A$, and $\gamma_A$ is a codimension two minimal area surface whose boundary $\de \gamma_A$ coincides with $\de A$. This is called the
holographic entanglement entropy. Intuitively, the entanglement entropy $S_A$ measures how much information is hidden inside $B$, when we divide the total space into two parts $A$ and $B$
(we will give a precise definition later).

Notice that the minimal area surfaces are more general than horizons of static black holes because in the former the trace of extrinsic curvature is required to vanish, while in the latter each component of extrinsic curvature should be vanishing. Later, this holographic entanglement entropy is covariantly generalized into the case where
the spacetime is Lorentzian, which can be time-dependent in general \cite{HRT}. See the earlier review articles \cite{EEReview} for a comprehensive review of
holographic entanglement entropy. Refer to \cite{SolR} for a detailed review on connections between the entanglement entropy and the entropy of black holes. Also a brief introduction of the holographic entanglement entropy is can be found in the review article \cite{McG} on the application of holography to condensed matter physics.

Historically, the entanglement entropy has originally been introduced to quantum field theories in the attempt to understand the microscopic origin of black hole entropy \cite{Thooft,Bombelli,Srednicki,SolR,SuUg,FPST,Ja,Solok}. The entanglement entropy has first been studied by using AdS/CFT setups with horizons in the pioneering works \cite{HMS,MBH}.

The purpose of this article is to review recent progresses on the holographic
entanglement entropy especially focusing on the gravitational dynamics such as black hole
formations. For example, a time evolution of a black hole
can be quantitatively measured as the evolution of holographic entanglement entropy.
Even though the definition of horizon entropy is ambiguous in time-dependent black hole
backgrounds depending on the choice of a time slice, the definition of the holographic entanglement entropy is unique \cite{HRT}. This is one of the remarkable advantages of the holographic entanglement entropy. We will explain more details of these later.

The entanglement entropy offers us an important observable when the spacetime $M$ has an additional boundary which intersects its boundary $\de A$. In this case, the non-gravitational
theory lives on a manifold with a boundary. In the context of AdS/CFT, such a situation occurs
when the conformal field theory (CFT) is defined on a manifold with a boundary, called the boundary conformal field theory (BCFT). Recently, the entanglement entropy has been computed in this
AdS/BCFT setup and has been shown to characterize the BCFT, as we will review later. In this way, the entanglement entropy is useful when we would like to characterize a gravitational spacetime with a non-trivial topology.

In the present article, we will only give a minimum guide to the entanglement entropy in quantum many-body systems, which nevertheless suffices to understand the rest of the material.
Refer to \cite{CaCaR,CaCaRR,CaHuR} for the review papers on entanglement entropy in quantum field theories and to \cite{RiLa,AFOV,ECP} for those in quantum many-body systems. We would also like to mention that the entanglement entropy has recently been applied to condensed matter physics as a new order parameter which classifies various quantum phases, though we will not discuss this aspect here.

This article is organized as follows: In section 2, we explain the definition and basic properties of entanglement entropy. Later we explain the holographic entanglement entropy based on the AdS/CFT. We will give a brief introduction to the AdS/CFT. In section 3, we discuss the holographic entanglement entropy in the presence of black holes. We also explain the analysis of black hole formations by employing the holographic entanglement entropy. In section 4, we explain the holographic dual of CFT on a manifold with a boundary. In section 5, we summarize conclusions and discuss future directions.

\section{Holographic Entanglement Entropy from AdS/CFT}

Here we first introduce the basic definition and properties of the
entanglement entropy in quantum many-body systems. After that we
will explain the holographic entanglement entropy with a brief
introduction to the AdS/CFT correspondence.

\subsection{Definition and Properties of Entanglement Entropy}

\begin{figure}[htbp]
   \begin{center}
     \includegraphics[height=5cm]{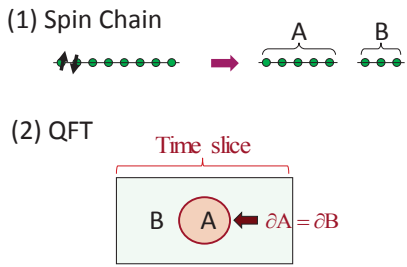}
   \end{center}
\caption {Examples of choices of subsystem $A$.}\label{subs}
\end{figure}

A state in quantum mechanics is described by a vector $|\Psi\lb$ in
a Hilbert space ${\cal H}$, which evolves in time by its Hamiltonian
$H$. Let us assume that the quantum system we consider has
multiple degrees of freedom (e.g. the quantum mechanics for more
than one particles) so that we can decompose the total system
into two subsystems $A$ and $B$. Accordingly, the total
Hilbert space ${\cal H}$ becomes a direct products \be {\cal
H}={\cal H}_A\otimes {\cal H}_B. \ee For example, we can consider a
spin chain, where a lot of spins are arrayed in a line as in Fig.\ref{subs}.
In this
example, we can choose $A$ and $B$ in many different ways just by
cutting the chain at an arbitrary point.

In quantum mechanics, physical quantities are computed as
expectation values of operators  as follows \be \la O \lb=\la\Psi
|O|\Psi\lb=\mbox{Tr}[\rho\cdot O], \label{denmat} \ee where we
defined the density matrix $\rho=|\Psi\lb \la \Psi|$. This system is
called a pure state as it is described by a unique wave function
$|\Psi\lb$. In more general cases, called mixed states, the system
is described by a density matrix $\rho$ as in (\ref{denmat}) instead
of the wave function $|\Psi\lb$, normalized such that Tr$\rho=1$. A
typical example of a mixed state is the canonical distribution
$\rho=e^{-\beta H}/\mbox{Tr}[e^{-\beta H}]$ at finite temperature
$T=\beta^{-1}$.

We define the reduced density matrix $\rho_A$ for the subsystem $A$
by tracing out with respect to ${\cal H}_B$ by \be
\rho_A=\mbox{Tr}_B[\rho]. \ee Then the entanglement entropy is
defined as the von-Neumann entropy for $\rho_A$ \be
S_A=-\mbox{Tr}[\rho_A\log \rho_A]. \ee

In the context of this paper, we consider the entanglement entropy
in quantum field theories. We can view a quantum field theory (QFT)
as an infinite copies of quantum mechanics. Therefore, its Hilbert
space ${\cal}$ is given by all possible field configurations of QFT
at a fixed time. Thus ${\cal H}_A$ is defined as those included in the
subspace $A$ on a fixed time slice\footnote{Recently, the entanglement
entropy for the subsystem $A$ by a region in the momentum
space has been analyzed in \cite{BMR}.}. In this way we can geometrically
define the subsystem $A$ as in Fig.\ref{subs}. The choice of $A$ is uniquely defined
in terms of the boundary $\de A$. There are obviously infinitely
different definitions of the entanglement entropy $S_A$ depending on the
choices of $A$.

We would like to summarize the basic properties of the entanglement entropy, which are useful
 in later arguments (refer to \cite{Nielsen-Chuang00} for more details).
 If the total system is a pure state, the equality
 $S_A=S_B$ is always satisfied. This means that the entanglement entropy for a pure state is not extensive as opposed to the thermal entropy. Also, for any systems, when we divide the system into four subsystems $A$, $B$, $C$ and $D$ so that there are no overlap between each of them
 i.e. ${\cal H}={\cal H_A}\otimes{\cal H_B}\otimes{\cal H_C}\otimes{\cal H_D}$, the following inequality is always satisfied \cite{LiRu}:
\be S_{A\cup B}+S_{B\cup C}\geq S_{A\cup B\cup C}+S_{B}. \label{SSA}
\ee This inequality is called the strong subadditivity
\cite{LiRu,Casiniarea}. These are properties which are true for any
quantum mechanical systems. It was shown in \cite{Casinicth} that we
can derive the c-theorem from the strong subadditivity if applied to
two dimensional relativistic field theories. Recently, an extension of
this analysis has been obtained in three dimensional field theories
by applying the strong subadditivity to infinitely many subsystems in \cite{Casini:2012ei},
which offers us an entropic proof of so called F-theorem \cite{Fth}.

One more useful property is the area law for quantum field theories.
Since the quantum field theories have infinitely many degrees of freedom,
the entanglement entropy $S_A$ is divergent. It has been shown that the leading divergence term is proportional to the area of the boundary $\de A$ \cite{Bombelli,Srednicki}:
\be
S_A=\gamma \cdot \f{\mbox{Area}}{a^{d-1}}+O(a^{-(d-2)}), \label{arealaw}
\ee
where $\gamma$ is a numerical constant; $a$ is the ultra-violet(UV) cut off in quantum field theories which is proportional to the lattice constant. The continuum limit corresponds to $a\to 0$. This is called the area law (see also e.g.\cite{Das,Casiniarea}). This has been proved for free field theories \cite{Eisert,RiLa,AFOV,ECP}. Even though for interacting field theories,
there has been no systematic direct test of area law, the holographic calculation using
the AdS/CFT \cite{RuTa,RuTaL} implies that the area law is true for any interacting quantum field theories which have UV fixed points. See also \cite{Ban} for a consistency condition for
the entanglement entropy in QFTs.

We should mention that there is an important exception of area law
(\ref{arealaw}). In two dimensional field theories, the area law is
violated in a logarithmic way if they are scale invariant. Since
scale invariant theories have a conformal symmetry, they are called
conformal field theories (CFTs). A simplest example of a CFT is a
free massless scalar field theory. In a two dimensional CFT defined
on a infinitely extended line, we have the following general result
\cite{HLW,Cardy} \be S_A=\f{c}{3}\log \f{l}{a},\label{tdcft} \ee
where $c$ is the central charge of the CFT; $l$ is the length of
subsystem $A$. In this way the $S_A$ has the logarithmic divergence.
For more results on the entanglement entropy in two dimensional
CFTs, refer to e.g.\cite{Cardy,CaCaR,CaCaRR,CaHuR}. A partial list
of the analysis of entanglement entropy in massive field theories or
higher dimensional field theories can be found in
\cite{Kabat,Casini05a,Fur,CaHuu,Shiba}.

\subsection{Holography and AdS/CFT}

The holographic principle argues that a gravitational theory in a
$d+2$ dimensional spacetime $M$ is equivalent to a non-gravitational
theory on a $d+1$ dimensional spacetime $\de M$, which is the
boundary of $M$ \cite{Hol}. The latter theory is typically described
by a quantum many-body system. A concrete example of holography is
known as the AdS/CFT correspondence \cite{Malda,GKPW}. The AdS/CFT
correspondence argues that a gravity on a $d+2$ dimensional Anti-de
Sitter space (AdS$_{d+2}$) is equivalent to a CFT on $d+1$
dimensional boundary of AdS$_{d+2}$, which is called
AdS$_{d+2}/$CFT$_{d+1}$.
 A typical choice of the
coordinate of AdS space is the Poincare coordinate, where the metric
of AdS$_{d+2}$ is given by \be ds^2=R^2\f{dz^2+dx^\mu dx_\mu}{z^2},
\label{AdSmet} \ee where $\mu=0,1,\ddd,d$. The parameter $R$ is
called the radius of AdS. In this case, the boundary of AdS$_{d+2}$
is given by the spacetime spanned by $(x^0,x^1,\ddd,x^d)$ at $z=0$.
Since the metric at $z=0$ gets divergent, we need to introduce the
cut off as $z>a$, using an infinitesimally small constant $a$. This
cut off in the AdS space is equivalent to the UV cut off $a$ in CFT
up to an order one constant. The important fact is that this new
coordinate $z$ corresponds to the length scale (or inverse of energy
scale) of the dual CFT in the sense of RG-flow.

The basic principle of AdS/CFT is called the bulk to boundary relation \cite{GKPW}.
This argues that the partition function of CFT is equal to that of the gravity on the AdS space i.e. $Z_{CFT}=Z_{AdS}$. In the classical gravity limit, which is assumed in the most parts of this article, the gravity partition function is just given in terms of the one-shell action
$I_{AdS}$ as $Z_{AdS}=e^{-I_{AdS}}$ in the Euclidean signature.

The AdS/CFT correspondence was originally found by considering near
horizon geometries of D-branes in string theory \cite{Malda}. Even
though we need to know the details of this in order to precisely
identify CFTs which is dual to the AdS spaces with various radius
$R$, we will not get into the details as they are not necessarily
crucial to understand the concept of AdS/CFT described below.
Instead, we would like to ask the readers to refer to string theory
literature e.g. the review \cite{adsreview} on this string theoretic
understandings. Here we would like to simply mention that the most
useful conclusion which can be obtained from the string theory
arguments can be summarized as follows. The dual CFTs are usually
given by $SU(N)$ Yang-Mills gauge theories with a ('t Hooft)
coupling constant $\lambda$, corresponding to $N$ D-branes. The
classical gravity limit (or called supergravity limit) is given by
the limit where both $N$ and $\lambda$ are taken to be infinitely
large. In this limit, the string theory is reduced to the
supergravity, which can be regarded as the general relativity
coupled to other fields such as the scalar fields and gauge fields.
This is because the large $N$ limit suppresses the quantum gravity
effect and the large coupling limit suppresses the string theoretic
corrections.

The AdS/CFT can be applied to more general background. We can modify the infrared (IR) geometry
i.e. the large $z$ region. We always require that in the boundary limit $z\to 0$, the metric
approaches that of the pure AdS (\ref{AdSmet}), which is called the asymptotically
AdS condition. Though it is believed that we can extend the AdS/CFT to more general backgrounds which are not asymptotically AdS, we will not discuss this here.

To understand the AdS/CFT better, it seems very important to study how the information in the CFT is encoded in that in the gravity theory. Especially, we can consider the information included in a certain region $A$ in the CFT and ask what is dual to it in the AdS gravity. Since the amount of information in the region $A$ can be measured by the entanglement entropy $A$, it will be interesting to consider what is the AdS dual of the entanglement entropy in a CFT. To answer this question is the main subject of this article.

\subsection{Holographic Entanglement Entropy}

\begin{figure}[htbp]
   \begin{center}
     \includegraphics[height=5cm]{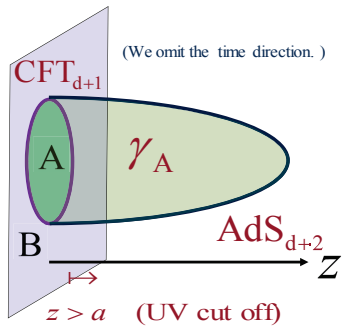}
   \end{center}
\caption {The calculation of holographic entanglement
entropy.}\label{holEE}
\end{figure}

In \cite{RuTa,RuTaL}, by applying the AdS/CFT correspondence, it is
argued that the entanglement entropy $S_A$ in a CFT can be
holographically calculated by the following formula of holographic
entanglement entropy (see Fig.\ref{holEE}): \be
S_A=\mbox{Min}_{\Sigma_A}\left[\f{\mbox{Area}(\Sigma_A)}{4G_N}\right],
\label{rt} \ee where $\Sigma_A$ is a codimension two surface (i.e.
$d$ dimensional in AdS$_{d+2}$) which satisfies $\de \Sigma_A=\de
A$; we also require that $\Sigma_A$ is homologous to $A$. The
minimum in (\ref{rt}) is taken for all surfaces $\Sigma_A$ which
satisfy this condition. Therefore $\Sigma_A$ is finally becomes the
minimal area surface $\gamma_A$ as in (\ref{HEE}). This formula
(\ref{rt}) can be applied to any static setups. The minimal area
surface is well-defined in the static case because we can
equivalently consider a Euclidean AdS space.

When the background is time-dependent, we need to employ the
covariant holographic entanglement entropy \cite{HRT}, which is
given by replacing $\Sigma_A$ with the extremal surface in the
Lorentzian asymptotic AdS space which satisfies the previous
condition. This covariant description corresponds to the
minimization of the Bousso's covariant entropy bound \cite{Bousso}.
If there are several extremal surfaces we take the one with minimum
area.

It is straightforward to see that the holographic entanglement entropy (\ref{rt}) leads to the area law as long as the gravity lives on an asymptotically AdS space, which is dual to a field theory with a UV fixed point. This is because the AdS metric gets divergent at $z=0$ and
this near boundary region gives the dominant divergent contributions to the area of minimal surface which is obviously proportional to the area of $\de A$.

\begin{figure}[htbp]
   \begin{center}
     \includegraphics[height=5cm]{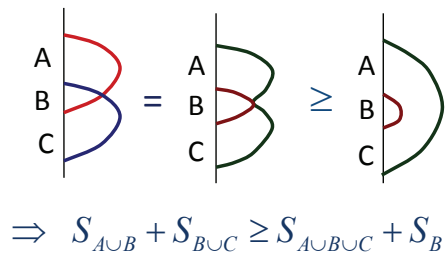}
   \end{center}
\caption {The holographic proof of strong
subadditivity. In each of three figures, the vertical black line represents the boundary of
the AdS, while the horizontal direction in the right is the $z$ direction in AdS. Though we are assuming the time slice of AdS$_3$ just for simplicity, this argument can be extended into higher dimension in a straightforward way. In the left picture, the red and blue curve represents the minimal surfaces $\gamma_{A\cup B}$ and $\gamma_{B\cup C}$. In the middle picture, we just recombine them into two surfaces (green and brown ones). The true minimal surfaces
$\gamma_{A\cup B\cup C}$ and $\gamma_{B}$ are given by the right picture. Therefore the strong subadditivity is obvious.}\label{HSSA}
\end{figure}

The strong subadditivity (\ref{SSA}) can also be holographically
proven very quickly for the holographic entanglement entropy in
static backgrounds \cite{Headrick:2007km} (see also \cite{HiTa}).
The essence of this proof is summarized in the Fig.\ref{HSSA}. This
only employs the fact that the holographic entanglement entropy is
given by a minimum of a certain integral on the surface $\Sigma_A$.
Moreover, it was recently found that another inequality called
monogamy can be derived in an analogous way \cite{Hayden:2011ag}.
This proves the Cadney-Linden-Winter inequality \cite{CLW}, which is
known to be independent from the strong subadditivity and is known
to be always satisfied for any quantum systems.

Just to satisfy the strong subadditivity and other inequalities, we
can replace the area function with other functionals which include
e.g. higher derivatives of curvatures. Indeed, this degrees of
freedom needs to be employed to find the holographic entanglement
entropy for gravity theories with higher derivative corrections such
as the Gauss-Bonnet gravity as we briefly explain later. However,
for the Einstein gravity coupled to any matter fields, we should
choose the area functional. In the presence of black hole horizon in
the AdS space, the minimal surface tends to wrap on the horizon.
Thus in order to be consistent with the Bekenstein-Hawking formula
(\ref{BHF}), we are naturally lead to the area formula (\ref{rt}).

\subsection{Evidences}

Even though the holographic entanglement entropy formula (\ref{rt})
has not been proven at present, there have been many supporting
evidences and have been no counterexamples. Though a heuristic
understanding of the formula (\ref{HEE}) was given in \cite{Fu},
this argument is not complete as pointed out in \cite{Matt} (see
also \cite{Casini:2012rn}). Thus, instead of giving a proof, below
we would like to list some of important evidences.

\begin{itemize}
  \item As we explained before, we can derive the area law and strong
  subadditivity from the holographic formula (\ref{HEE}).\\

  \item We can explicitly confirm that in the AdS$_3$/CFT$_2$, the holographic
  entanglement entropy precisely agrees with the CFT
  result\cite{RuTa,RuTaL,Mic}. This can be seen as follows.
We start with the Poincare metric of AdS3 \be
ds^2=R^2\f{dz^2-dt^2+dx^2}{z^2}. \label{adsthr}\ee The two
dimensional CFT lives on the space spanned by $t$ and $x$. We choose
the subsystem $A$ to be the length $l$ interval $|x|\leq l/2$ in the
infinitely long total space $-\infty<x<\infty$. In AdS$_3$/CFT$_2$,
the minimal surface $\gamma_A$ is given by a geodesic line in
AdS$_3$ on a timce slice $t=$constant. It is an elementary exercise
to see that it is given by a half circle with radius $l/2$ i.e.
$x=\s{l^2/4-z^2}$. Since the induced metric on this geodesic is
given by \be ds^2_{ind}=\f{l^2 dz^2}{4z^2\s{l^2/4-z^2}}. \ee The
holographic entanglement entropy now reads as \be
S_A=\f{R}{2G_N}\int^{l/2}_a dz
\f{l}{2z\s{l^2/4-z^2}}=\f{R}{2G_N}\log \f{l}{a}= \f{c}{3}\log
\f{l}{a}.\ee Here we employed the relation $c=\f{3R}{2G_N}$ between
the central charge of 2 dim. CFT and the radius of AdS$_3$
\cite{BH}. This precisely agrees with the CFT result
(\ref{tdcft}).\\

  \item The proof of holographic formula (\ref{HEE}) in the special
  case where $A$ is a round sphere has been given in \cite{CHM} for any dimensions.
  This analysis has been generalized to calculate the Renyi entropy \cite{HungR} (see
also\cite{KPSS,FurR}).\\

  \item  In the setup of AdS$_{d+2}/$CFT$_{d+1}$, for a generic,
  smooth and compact subsystem $A$, the holographic entanglement
  entropy behaves as follows \cite{RuTa,RuTaL}:
\begin{eqnarray}
S_{A} &=&
  p_1  \left(l/a\right)^{d-1}
+ p_3 \left(l/a \right)^{d-3}
+\cdots  \label{areatwo}  \\
%\hphantom{AAAAAAAAAAA}
&&
%\\%%%%%
%\quad
\cdots +\left\{
\begin{array}{ll}
\displaystyle p_{d-1}\left(l/a\right) + p_d + \mathcal{O}(a/l) \ , &
 \mbox{$d+1$: odd} \ ,   \\
\displaystyle p_{d-2} \left(l/a\right)^{2} + q \log
\left(l/a\right)+ \mathcal{O}(1) \ ,
&  \mbox{$d+1$: even} \ .   \\
\end{array}
\right.
%&&
 \nonumber
\end{eqnarray}

Thus there is a logarithmic divergent term in even dimensional CFTs,
This is a universal term in that its coefficient $q$ is independent
from the UV cut off $a$. In general, $q$ is proportional to a linear
combination of the central charges in CFT$_{d+1}$. We already
explicitly explained that this agrees with the CFT$_2$ result. In
the higher dimensional cases, the agreement of $q$ between the
AdS$_{d+2}$ and CFT$_{d+1}$ has been confirmed in
\cite{RuTaL,Sod,MySi,CHM} (see also \cite{ScTh,Paulos:2011zu,Fuj}).

On the other hand, in odd dimensional CFTs, we find that the finite
constant $p_d$ is independent from the cut off. It has been
suggested that $p_d$ can be used as a measure of degrees of freedom
in odd dimensional CFT, where there are no conformal anomalies and
central charges \cite{MySi,CHM,LiMe}. This is expected to be related
to the F-theorem in three dimensional CFTs \cite{Fth}. See also
\cite{Albash:2011nq, Myers:2012ed} for other relations between the
entanglement entropy and RG flow. Refer to
\cite{LNSS,CaHuS,Dowa,Solrr,Dowb,Dowc} for calculations of $q$ and
$p_d$ in free field theories. See also \cite{MFS,KPSSe} for the
analysis of the entanglement entropy in three dimensional $O(N)$
vector models and Chern-Simons theories.

The holographic entanglement entropy was recently analyzed in the
presence of relevant perturbation in \cite{HungC}, where extra
logarithmic contributions have been observed. A similar result has
already been obtained in the free scalar field theory\cite{Wil}.\\

\item If we consider the holographic entanglement entropy in gravity duals of
confining gauge theories such as the AdS soliton \cite{Witten} and
Klebanov-Strassler solutions \cite{KlSt}, we find that the
derivative of $S_A$ with respect to the size of $A$ gets
discontinuous at some point \cite{NiTa,KKM,Pak,OgTa,Ishihara:2012jg,Cai:2012sk,Lewkowycz:2012mw}. This is
considered to be dual to the confinement/deconfinement phase
transition dual to the Hawking-Page transition \cite{HaPe}. The
lattice calculations \cite{BuPo,Nak} (see also \cite{Ve}) of pure
Yang-Mills theory qualitatively confirm this prediction from
AdS/CFT, though the order of phase transition is no longer first
order for these finite $N$ calculations. In particular, it was shown
that the holographic entanglement entropy computed for the AdS
soliton geometry precisely agrees with that computed in the free
field theory \cite{NiTa} when supersymmetry is only weakly broken.
An analogous result is obtained for the geometric entropy in
\cite{FNT,FO}. For studies of holographic entanglement entropy for
some other gravity duals including non-conformal theories
refer to \cite{FreedBrown:2009py,Arean:2008az,FLRT,Ramallo:2008ew,ALT,Nie,AlJoh}.
It is possible that the entanglement entropy can be a useful
probe of QCD \cite{Cornwall:2012si}.\\

  \item Consider the case where the subsystem $A$ consists of
  disconnected regions e.g. $A_{1}\cup A_{2}$.
  The holographic entanglement entropy predicts phase transitions
when we change the distance between $A_1$ and $A_2$
\cite{Matt,Tonn}. In the 2 dim. CFT, these results have been shown
to be consistent with those in CFT \cite{Matt} in a non-trivial way.
For relevant calculations in the CFT side
refer to \cite{FPS,CaHuD,CCT,RaGl}.\\
\end{itemize}

\subsection{Higher Derivative Corrections}

The holographic formula (\ref{HEE}) assumes the classical gravity
limit of string theory, which corresponds to the large $N$ and
strongly coupled limit of dual gauge theories. Therefore it is very
intriguing to see how this formula is modified in the presence of
corrections. In string theory, there are two quantum corrections:
one is the quantum gravity corrections and the other is the stringy
corrections as we mentioned. At present, we have little
understanding on the former and thus here we will concentrate on the
stringy corrections. These are described by higher derivative
corrections to the Einstein gravity. Even for them the understanding
is currently limited, the holographic entanglement entropy has been
found only for the Lovelock
gravities\cite{Hung:2011xb,deBoer:2011wk} (see also later
developments \cite{OgTa,Yale:2011dq}). Let us briefly review this in
the simplest example: Gauss-Bonnet gravity. Its gravity action looks
like \be S_{GB}=-\f{1}{16G_N}\int
dx^{d+2}\s{g}\left[R-2\Lambda+\lambda
(R_{\mu\nu\rho\sigma}R^{\mu\nu\rho\sigma}-4R_{\mu\nu}R^{\mu\nu}+R^2)\right],
\ee where $\Lambda<0$ is the negative cosmological constant of AdS
space and $\lambda$ is the Gauss-Bonnet parameter. The holographic
entanglement entropy is argued to be
\cite{Hung:2011xb,deBoer:2011wk} \be S_A=
\mbox{Min}_{\Sigma_A}\left[\f{1}{4G_N}\int_{\Sigma_A}dx^d\s{h}(1+2\lambda
R_{int})\right], \label{holde} \ee where $R_{int}$ is the intrinsic
curvature of $\Sigma_{A}$. This formula passes several non-trivial
tests. See also \cite{Fu,Hyun:2012mh} for other aspects of higher
derivative corrections to the holographic entanglement entropy.

\section{Black Hole Formations and Quantum Quenches}

So far we discussed the AdS/CFT at zero temperature or equally at a ground state.
If we heat up the system, the CFT reaches to a thermal equilibrium state at a finite temperature. This finite temperature CFT is dual to a black hole in the AdS space \cite{Witten}. The AdS/CFT in this case nicely fits with the well-know fact that the black hole follows thermodynamics. Moreover if we consider the process of the heating up the system, where the temperature getting increasing, the gravity dual is described by a black hole formation in the AdS space. It is quite remarkable that the AdS/CFT allows us to analyze strongly coupled non-equilibrium systems. Below we would like to discuss the behavior of the holographic entanglement in these situations.

\subsection{Holographic Entanglement Entropy at Finite Temperature}

Consider a calculation of the holographic entanglement entropy
at finite temperature $T=\beta^{-1}$ in the simplest example of AdS/CFT i.e. the AdS$_3/$CFT$_2$ duality. We assume that
the spatial length of the total system $L$ is infinite i.e.
$\beta/L\ll 1$. In such a high temperature region, the gravity
dual of the conformal field theory is described by the Euclidean BTZ
black hole \cite{BTZ}. Its metric looks like \be
ds^2=(r^2-r_+^2)d\tau^2+ \f{R^2}{r^2-r^2_{+}}dr^2+r^2 d\vp^2 \ .
\label{btzmet}\ee Note that if we set $z=1/r$ and perform trivial coordinate rescalings,
we can confirm that this metric approaches to the pure AdS$_3$ (\ref{adsthr}) in the $r\to \infty$
limit.

The Euclidean time is compactified as
$\tau\sim\tau+\f{2\pi R}{r_+}$ to obtain a smooth geometry. We also
impose the periodicity $\vp\sim \vp+2\pi$. By taking the boundary
limit $r\to \infty$, we find the relation between the boundary CFT
and the geometry (\ref{btzmet}) \be \f{\beta}{L}=\f{R}{r_{+}}\ll 1 \ .
\label{relationbtz}\ee

The subsystem for which we consider the entanglement entropy is
given by $0\leq \vp\leq 2\pi l/L$ at the boundary. Then by extending
our formula (\ref{arealaw}) to asymptotically AdS spaces,
 the entropy can be computed from the
length of the space-like geodesic starting from $\vp=0$ and ending
at $\vp=2\pi l/L$ at the boundary $r=r_0\to \infty$ at a fixed time.
This geodesic distance can be found analytically as
\be \cosh\left(\f{\mbox{Length}(\gamma_A)}{R}\right)
=1+\f{2r_0^2}{r_+^2}\sinh^2\left(\f{\pi
l}{\beta}\right) \ .\label{delltwo} \ee
The relation between the cut off $a$ in CFT and
the one $r_0$ of AdS is given by $\f{r_0}{r_+}=\f{\beta}{a}$.
Then it is easy to see that our area law
(\ref{arealaw}) precisely reproduces the known CFT result \cite{RuTa,RuTaL}
given by the following formula \cite{Cardy,CaCaR,CaCaRR}
\be
S_A=\f{c}{3}\log
\left(\f{\beta}{\pi a}\sinh\left(\f{\pi l}{\beta}\right)\right).
\label{entropytemp}
\ee

\begin{figure}[htbp]
   \begin{center}
     \includegraphics[height=7cm]{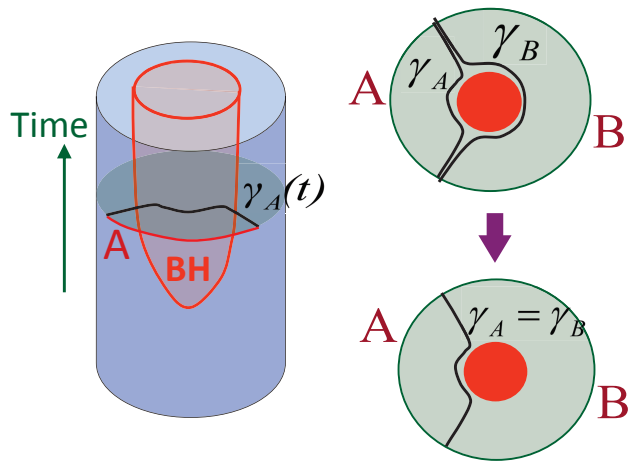}
   \end{center}
\caption {The left figure schematically describes the black hole creation and the extremal surface $\gamma_A$. The orange curve represents the time-evolving black hole. The right figures describe how the extremal surface $\gamma_A$ and $\gamma_B$ should be chosen. In the black hole creation spacetime, the right lower picture describes the correct choice i.e. $\gamma_A$ and $\gamma_B$ coincides. On the other hand, for eternal blackholes (i.e time-independent black holes), we need to distinguish $\gamma_A$ and $\gamma_B$ as in the right upper figure.}\label{fig:quench}
\end{figure}

It is also useful to understand these calculations geometrically.
The geodesic line in the BTZ black hole takes the form shown in
the right upper picture in Fig.\ \ref{fig:quench}. When the size of $A$ is
small, it is almost the same as the one in the ordinary AdS$_3$. As
the size becomes large, the turning point approaches the horizon and
eventually, the geodesic line covers a part of the horizon. This is
the reason why we find a thermal extensive behavior of the entropy when
$l/\beta\gg 1$ in (\ref{entropytemp}). The thermal entropy in a
conformal field theory is dual to the black hole entropy in its
gravity description via the AdS/CFT correspondence. In the presence
of a horizon, it is clear that $S_A$ is not equal to $S_B$ (remember
$B$ is the complement of $A$) since the corresponding geodesic lines
wrap different parts of the horizon
(see the right upper picture in Fig.\ \ref{fig:quench}).
This is a typical property of the entanglement entropy for a mixed state and
thus the topological obstruction due to the black hole horizon directly corresponds to
the basic property of mixed states.

\subsection{Holographic Entanglement Entropy and Black Hole Formations}

\begin{figure}[htbp]
   \begin{center}
     \includegraphics[height=5cm]{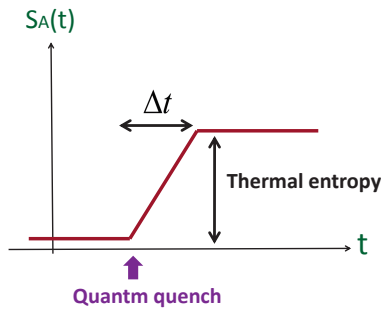}
   \end{center}
\caption {The time evolution of the entanglement entropy in a two dimensional CFT.}\label{fig:time}
\end{figure}

A more interesting backgrounds in AdS/CFT is time-dependent solutions where a black hole is
formed \cite{HLR,ChYa,Bhattacharyya:2009uu,DNT,HR}. A simple class of such examples of time-dependent backgrounds in QFTs is called quantum quenches \cite{CCa,CCb,AsAv}. A quantum quench is triggered by a sudden shift of parameters such as the mass in a quantum field theory. This means that the injection of energy is taken place instantly, shifting a ground state into an excited state at a given time. If the theory at later time is massless, we can regard the system as an exited state in a CFT.  One of the important quantities
which characterize such a time evolution is the entanglement entropy. As shown in \cite{CCa},
the entanglement entropy under a quantum quench in two dimensional CFTs always increases linearly as a function of time and eventually reaches a constant value after the thermalization time
$\Delta t$ as sketched in Fig.\ref{fig:time}. The increased amount of the entanglement entropy at late time $t>\Delta t$ is the same as the thermal entropy at the final thermal equilibrium. The thermalization time $\Delta t$ is found to be a half of the length $l$ of the subsystem $A$ \cite{CCa}, which is explained by assuming that the information propagates at a speed of light.

Analysis of the time evolution of entanglement entropy has started
in \cite{HRT} by employing the covariant holographic entanglement
entropy, where the entangling surface $\gamma_A$ is given by an
extremal surface in the AdS space. Recently, there have been
remarkable developments on studies of time-dependent holographic
entanglement entropy
\cite{AAL,AC,TaUg,Balth,Hubi,BlasMu,FMS,AlTo,MuSc,KKTh,Hub,Local,CHH}.
For example, it has been confirmed that the holographic analysis
agrees with the two dimensional CFT result \cite{AAL}. In higher
dimensional CFTs, the holographic analysis reveals that the
thermalization time $\Delta t$ depends on the shape of the subsystem
$A$ \cite{AC,Balth,Hub}. When $A$ is a round ball with the radius
$l$, then we have $\Delta t=\f{l}{2}$, while when $A$ is an
infinitely extended strip with the width $l$ we find $\Delta
t>\f{l}{2}$. The consistency with the strong subadditivity has been
confirmed for various time-dependent examples in \cite{AlTo,CHH}. Interesting
oscillating modes have been found in \cite{FMS}. The time evolution
of holographic entanglement entropy under local quenches has been
studied in \cite{Local}.

Moreover, the holographic entanglement entropy allows us to answer
the basic puzzle on the entropy of time-dependent back holes
\cite{AAL,TaUg}. In a thermalization of CFT, an initially pure state
gets excited and evolves until it reaches the thermal equilibrium.
Its gravity dual is a black hole formation in the AdS space. At
early time, the spacetime is the pure AdS, while at late time, it
approaches a static AdS black hole. Thus one might be tempting to
conclude that the entropy, which is initially vanishing, should
increase under the time evolution. This clearly contradicts with its
CFT dual, where a pure state should follow a unitary evolution,
which does not change the microscopic entropy. The total entropy
$S_{tot}=-\mbox{Tr}\rho\log \rho$ is conveniently calculated from
the difference of the entanglement entropy \be S_{tot}=\lim_{|B|\to
0} \left(S_A-S_B\right). \label{forab} \ee Indeed, for a pure state,
which always satisfies $S_A=S_B$, we find $S_{tot}=0$ by using this
formula. Then the holographic analysis explained in the
Fig.\ref{fig:quench} shows that the total entropy is actually
vanishing at any time. In this case, the presence of horizon is not
a topological obstruction as we can modify the surfaces $\gamma_A$
and $\gamma_B$ so that it topologically equivalent. In this way, we
can conclude that during a black hole formulation, the microscopic
entropy does not increase \cite{AAL,TaUg}. However, the
coarse-grained entropy is increasing as its apparent horizon
expands. Actually, we can regard the entanglement entropy $S_A$ as a
coarse-grained entropy because we trace out some part of the space
and $S_A$ indeed increases under the time evolution by the amount of
the thermal entropy in the final equilibrium. Notice that there is
no obvious unique definition for the entropy of a time-dependent
black holes, while the definition of holographic entanglement
entropy is unique even for time-dependent backgrounds.

\section{Holographic Dual of BCFT}

\begin{figure}[htbp]
   \begin{center}
     \includegraphics[height=5cm]{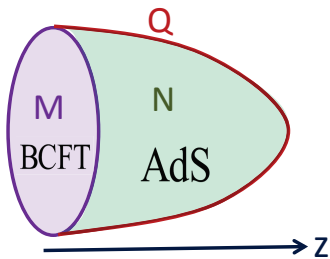}
   \end{center}
\caption {A sketch of general setups of holographic dual of BCFT}\label{Fig:set}
\end{figure}

As a final topic we would like to consider the holographic dual of
CFT defined on a manifold $M$ with a boundary $\de M$, which is so
called boundary conformal field theory (BCFT). This is argued to be
given by generalizing the AdS/CFT correspondence in the following
way \cite{Takayanagi:2011zk,Fujita:2011fp} (called as AdS/BCFT).
Based on the idea of holography and AdS/CFT, we extend a $d+1$
dimensional manifold $M$ to a $d+2$ dimensional asymptotically AdS
space $N$ so that $\de N=M\cup Q$, where $Q$ is a $d+1$ dimensional
manifold which satisfies $\de Q=\de M$. See Fig.\ref{Fig:set} for
this setup.

Usually, we impose the Dirichlet boundary condition on the metric at
the boundary of AdS and following this we assume the Dirichlet
boundary condition on $M$. On the other hand, we require
a Neumann boundary condition on the metric at $Q$, whose details
will be explained later. This change of boundary condition is the most important
part of the holographic construction of BCFT.

In specific setups, such a holography construction of BCFT has already been mentioned in the earlier papers \cite{Kaa,Kab}. Different constructions of holographic dual of field theories with boundaries can be found in \cite{AMR,ABBS,GutB}. Moreover, our setup can be regarded as a modification of the well-known Randall-Sundrum setup \cite{RaSu} such that the additional boundary $Q$ intersects with the original asymptotically AdS boundary.

\subsection{Construction}

To make the variational problem sensible, we add the
Gibbons-Hawking boundary term \cite{GHterm} to the Einstein-Hilbert
action (we omit the boundary term for $M$): \be I=\f{1}{16\pi
G_N}\int_{N}\s{-g}(R-2\Lambda)+\f{1}{8\pi G_N}\int_{Q}\s{-h}K.
\label{Loein} \ee The metric of $N$ and $Q$ are denoted by $g$ and
$h$, respectively. $K=h^{ab}K_{ab}$ is the trace of extrinsic
curvature $K_{ab}$ defined by $K_{ab}=\nabla_a n_b,$ where $n$ is
the unit vector normal to $Q$ with a projection of indices onto $Q$
from $N$.

Consider the variation of metric in the above action. After a partial integration, we find
\be
\delta I=\f{1}{16\pi G_N}\int_{Q}\s{-h}\left(K_{ab}\delta h^{ab}-Kh_{ab}\delta h^{ab}\right).
\ee
Notice that the terms which involve the derivative of $\delta h_{ab}$ cancels out
thanks to the boundary term.
We can add to (\ref{Loein}) the action $I_Q$ of some matter fields localized on $Q$.
We impose the Neumann boundary condition instead of the Dirichlet one by setting
the coefficients of $\delta h^{ab}$ to zero and finally
we obtain the boundary condition
\be
K_{ab}-h_{ab}K=8\pi G_N T^{Q}_{ab}, \label{bein}
\ee
where we defined
\be
T^{Qab}=\f{2}{\s{-h}}\f{\delta I_Q }{\delta h_{ab}}. \label{matbc}
\ee

As a simple example we would like to assume that the boundary matter
lagrangian is just a constant. This leads us to consider the
following action \be I=\f{1}{16\pi
G_N}\int_{N}\s{-g}(R-2\Lambda)+\f{1}{8\pi
G_N}\int_{Q}\s{-h}(K-T).\label{act} \ee The constant $T$ is
interpreted as the tension of the boundary surface $Q$. In AdS/CFT,
a $d+2$ dimensional AdS space (AdS$_{d+2}$) is dual to a $d+1$
dimensional CFT. The geometrical $SO(2,d+1)$ symmetry of AdS is
equivalent to the conformal symmetry of the CFT. When we put a $d$
dimensional boundary to a $d+1$ dimensional CFT such that the
presence of the boundary breaks $SO(2,d+1)$ into $SO(2,d)$, this is
called a boundary conformal field theory (BCFT) \cite{Cbcft}. Note
that though the holographic duals of defect or interface CFTs
\cite{Kaa,Kab,Janus} look very similar with respect to the
symmetries, their gravity duals are different from ours because they
do not have extra boundaries like $Q$.

To realize this structure of symmetries, we take the following
ansatz of the metric (see also \cite{Kaa,Kab,BE}): \be
ds^2=d\rho^2+\cosh^2\f{\rho}{R}\cdot ds^2_{AdS_{d+1}}.\label{metads}
\ee If we assume that $\rho$ takes all values from $-\infty$ to
$\infty$, then (\ref{metads}) is equivalent to the AdS$_{d+2}$. To
see this, let us assume the Poincare metric of AdS$_{d+1}$ by
setting \be
ds_{AdS_{d+1}}^2=R^2\f{-dt^2+dy^2+d\vec{w}^2}{y^2},\label{metadss}
\ee where $\vec{w}\in R^{d-1}$. Remember that the cosmological
constant $\Lambda$ is related to the AdS radius $R$ by
$\Lambda=-\f{(d+1)d}{2R^2}$.

By defining new coordinates $z$ and $x$ as \be
z=y/\cosh\f{\rho}{R},\ \ x=y\tanh\f{\rho}{R}, \ee we recover the
familiar form of the Poincare metric of AdS$_{d+2}$:
$ds^2=R^2(dz^2-dt^2+dx^2+d\vec{w}^2)/z^2$.

To realize a gravity dual of BCFT, we will put the boundary $Q$ at
$\rho=\rho_*$ and this means that we restrict the spacetime to the
region $-\infty<\rho<\rho_*$. The extrinsic curvature on $Q$ reads
\be K_{ab}=\f{1}{R}\tanh\left(\f{\rho}{R}\right)h_{ab}. \ee The
boundary condition (\ref{bein}) leads to \be
K_{ab}=(K-T)h_{ab}.\label{eqbein} \ee Thus $\rho_*$ is determined by
the tension $T$ as follows \be
T=\f{d}{R}\tanh\f{\rho_*}{R}.\label{tension} \ee

\subsection{Boundary Entropy}

Let us concentrate on the $d=1$ case to describe the two dimensional
BCFT. This setup is special in that it has been well-studied (see
\cite{Cardyb} and references therein) and that the BCFT has an
interesting quantity called the boundary entropy (or $g$-function)
\cite{AfLu}. We define the quantity called $g$ by the partition
function on a disk denoted by $g_\ap$, where $\ap$ parameterizes the
choice of boundary conditions. The boundary entropy
$S^{(\ap)}_{bdy}$ is defined by \be S^{(\ap)}_{bdy}=\log g_{\ap}.
\label{gfun} \ee The boundary entropy measures the boundary degrees
of freedom and can be regarded as a boundary analogue of the central
charge $c$.

%\subsection{Boundary Entropy from Disk Partition Function}

Consider a holographic dual of a CFT on a round disk defined by
$\tau^2+x^2\leq r^2_D$ in the Euclidean AdS$_3$ spacetime \be
ds^2=R^2\f{dz^2+d\tau^2+dx^2}{z^2},\label{poth} \ee where $\tau$ is
the Euclidean time. In the Euclidean formulation, the action
(\ref{act}) is now replaced by \be I_E=-\f{1}{16\pi
G_N}\int_{N}\s{g}(R-2\Lambda)-\f{1}{8\pi
G_N}\int_{Q}\s{h}(K-T).\label{acte} \ee Note that $\rho_*$ is
related to the tension $T$ of the boundary via (\ref{tension}). When
the BCFT is defined on the half space $x<0$, its gravity dual has
been found in previous subsection. Therefore we can find the gravity
dual of the BCFT on the round disk by applying the conformal map
(see e.g.\cite{BM}). The final answer is the following domain in
AdS$_3$ \be
\tau^2+x^2+\left(z-\sinh(\rho_*/R)r_D\right)^2-r_D^2\cosh^2(\rho_*/R)\leq
0. \label{diskh} \ee In this way we found that the holographic dual
of BCFT on a round disk is given by a part of the two dimensional
round sphere. A larger value of tension corresponds to the larger
radius.

Now we would like to calculate the disk partition function in order to obtain
the boundary entropy. By evaluating (\ref{acte}) in the domain (\ref{diskh}), we obtain
\ba
\!\!I_E\!=\!\f{R}{4G_N}\!\!\left(\!\f{r^2_D}{2a^2}\!+\!\f{r_D\sinh(\rho_*/R)}{a}\!
+\!\log(a/r_D)\!-\!\f{1}{2}
\!-\!\f{\rho_*}{R}\!\!\right)\!\!,\ \ \ \ \label{diskpar}
\ea
where we introduced the UV cutoff $z>a$ as before.
By adding the counter term on the AdS boundary \cite{Ren},
we can subtract the divergent terms in (\ref{diskpar}). The difference of the partition function
between $\rho=0$ and $\rho=\rho_*$ is given by $I_E(\rho_*)-I_E(0)=-\f{\rho_*}{4G_N}$.
Since the partition function is given by $Z=e^{-I_E}$, we obtain the boundary entropy
\be
S_{bdy}=\f{\rho_*}{4G_N}, \label{bed}
\ee
where we assumed $S_{bdy}=0$ for $T=0$ because the boundary contributions
vanish in this case.

%\subsection{Boundary Entropy from Holographic Entanglement Entropy}

Another way to extract the boundary entropy is to calculate the
entanglement entropy. In a two dimensional CFT on a half line, $S_A$
behaves as follows \cite{Cardy,CaCaR,CaCaRR} \be
S_A=\f{c}{6}\log\f{l}{a}+\log g,\label{entg} \ee where $c$ is the
central charge and $a$ is the UV cut off (or lattice spacing); $A$
is chosen to be an interval with length $l$ such that it ends at the
boundary. The $\log g$ in (\ref{entg}) coincides with the boundary
entropy (\ref{gfun}).

In AdS/CFT, the holographic entanglement entropy can be calculated by the formula
(\ref{HEE}). Consider the gravity dual of a two dimensional BCFT on a half line
$x<0$ in the coordinate (\ref{poth}). By taking the time slice
$\tau=0$, we define the subsystem $A$ by the interval $-l\leq x\leq
0$. In this case, the minimal surface (or geodesic line) $\gamma_A$
is given by  $x^2+z^2=L^2$. If we go back to the coordinate system
(\ref{metads}) and (\ref{metadss}), then $\gamma_A$ is simply given
by $\tau=0, y=l$ and $-\infty<\rho\leq\rho_*$. This leads to \be
S_A=\f{1}{4G_N}\int^{\rho_*}_{-\infty}d\rho. \ee By subtracting the
bulk contribution which is divergent as in (\ref{entg}),
we reproduce the previous result (\ref{bed}).
See also \cite{GutB} for the recent calculation of boundary entropy in supergravity.
A similar calculations of boundary entropy for interface CFTs can be found in \cite{Gut}.

\subsection{Holographic g-theorem}

In two dimension, the central charge $c$  is the most important
quantity which characterizes the degrees of freedom of CFT.
Moreover, there is a well-known fact, so called c-theorem
\cite{cth}, that the central charge monotonically decreases under
the RG flow. In the case of BCFT, an analogous quantity is actually
known to be the g-function or equally boundary entropy \cite{AfLu}.
At fixed points of boundary RG flows, it is reduced to that of BCFT
introduced in (\ref{gfun}). It has been conjectured that the
g-function monotonically decreases under the boundary RG flow in
\cite{AfLu} and this has been proven in \cite{FrKo} later.
 Therefore the holographic proof of g-theorem described below
will offer us an important evidence of our proposed holography.
Refer to \cite{hcth} for a holographic c-theorem and to \cite{Ya}
for a holographic g-theorem in the defect CFT under a probe
approximation.

Because we want to keep the bulk conformal invariance and we know
that all solutions to the vacuum Einstein equation with $\Lambda<0$
are locally AdS$_3$, we expect that the bulk spacetime remains to be
AdS$_3$. We describe the boundary $Q$ by the curve $x=x(z)$ in the
metric (\ref{poth}). We assume generic matter fields on $Q$ and this
leads to the energy stress tensor $T^Q_{ab}$ term in the boundary
condition (\ref{bein}). It is easy to check the energy conservation
$\nabla^a T^{Q}_{ab}=0$ in our setup because
$\nabla^a(K_{ab}-Kh_{ab})=R_{n b}$, where $n$ is the Gaussian normal
coordinate which is normal to $Q$. In order to require that the
matter fields on the boundary are physically sensible, we impose the
null energy condition (or weaker energy condition) as in the
holographic c-theorem \cite{hcth}. It is given by the following
inequality for any null vector $N^a$ \be T^{Q}_{ab}N^aN^b\geq 0.
\label{nulle} \ee In our case, we can choose \be
(N^t,N^z,N^x)=\left(\pm
1,\f{1}{\s{1+(x'(z))^2}},\f{x'(z)}{\s{1+(x'(z))^2}}\right). \ee Then
the condition (\ref{nulle}) is equivalent to \be x''(z)\leq
0.\label{cond} \ee

Since at a fixed point the boundary entropy is given by
$S_{bdy}=\f{\rho_*}{4G_N}$ and we have the relation
$\f{x}{z}=\sinh(\rho_*/R)$ on the boundary $Q$, we would like to
propose the following $g$-function \be \log g(z)=\f{R}{4G_N}\cdot
\mbox{arcsinh}\left(\f{x(z)}{z}\right). \ee By taking derivative, we
get \be \f{\de \log g(z)}{\de z}=\f{x'(z)z-x(z)}{\s{z^2+x(z)^2}}.
\ee Indeed we can see that $x'z-x$ is non-positive because this is
vanishing at $z=0$ and (\ref{cond}) leads to $(x'z-x)'=x''z\leq 0$.
Thus we can show that $g(z)$ is a monotonically decreasing function
of $z$, which is dual to the length scale of the dual BCFT. In this
way, we manage to derive the g-theorem in our setup. We can
generalize this argument into higher dimensions
\cite{Fujita:2011fp}, which leads to a proposal of a higher
dimensional analogue of the g-theorem.  Refer to
\cite{AlFa,SeKa,AnUh,Kw,Fujita:2012fp} for other aspects of
AdS/BCFT.

\section{Conclusions}

In this review article, we presented a quick survey on the recent
progresses on holographic entanglement entropy (HEE). We can think
of several applications of HEE to various subjects. One of them will
be quantum mechanical understandings of black holes, which was
historically the original motivation of considering the entanglement
entropy in quantum field theories. For example, we explained that
the HEE can give a useful order parameter for black hole formation
processes which are dual to the thermalization of strongly coupled
systems. This is expected to give a nice relation between the black
hole physics and non-equilibrium physics. Refer also to
\cite{BEY,Cad,Aze,BaFa,DSS,CJP,Strominger:2009aj,Cadn,Swing,Sen,BarF,Sim,Tet,Cantc}
for other progresses on the HEE for black holes which we could not
discuss in the main context of this article. For applications of
holographic entanglement entropy to brane-world setups refer to
\cite{Em,Iw,SoB} (see also \cite{LiPa}), which was pioneered by
\cite{HMS}. Also, there have been studies of holography in
non-trivial spacetimes such as flat space \cite{BaFu,LiTa} and AdS
wormholes \cite{ABS,FuHaTa}.

The applications of HEE to condensed matter physics is also very intriguing.
For example, the HEE is employed to search gravity duals with Fermi surfaces \cite{OTU,HuSS,Shag,DHKTW,Kim:2012nb,Sachdev:2012dq}. An interesting behavior analogous to the entanglement entropy has been observed in a problem of image compression \cite{Matsu}.
It has been pointed out that the holographic entanglement entropy supports the
idea of emergent gravity in \cite{Raa} (see also \cite{Fup}). One
way to make this idea concrete seems to employ the conjectured
connection \cite{SwER} between the AdS/CFT and the multi scale
entanglement renormalization (MERA) \cite{ERn} (see also
\cite{MoSo}).

Then it is natural to ask how the quantum information in CFT is encoded in the AdS spacetime. It is well expected that the entanglement entropy will play an important role again here. The methods to extract the bulk metric from then holographic entanglement entropy
has been discussed in \cite{Ham,Bil,Hub}. Moreover, quite recently, there have been interesting discussions on reconstructions of the bulk geometry from the information on a certain region at the boundary
\cite{Bousso:2012sj,Czech:2012bh,Hubeny:2012wa}. Indeed, the idea of holographic entanglement entropy has turned out to be closely related \cite{Czech:2012bh,Hubeny:2012wa} and more detailed analysis certainly deserves a future study.

In this way, the entanglement entropy connects directly between
gravity backgrounds and quantum states in quantum many-body systems.
Though the metric in gravity may not be a good quantity to look at
in the presence of significant quantum corrections, the
(holographic) entanglement entropy should be well-defined even at
the quantum level. Thus the HEE should be useful for the
understanding of both quantum gravity and condensed matter systems.

\vspace{1cm}

{\bf Acknowledgements} TT would like to thank Tastuo Azeyanagi,
Sumit Das, Mitsutoshi Fujita, Tasuyuki Hatsuda, Matthew Headrick,
Tomoyoshi Hirata, Veronika Hubeny, Andreas Karch, Wei Li, Tatsuma
Nishioka, Noriaki Ogawa, Mukund Rangamani, Shinsei Ryu, Ethan
Thompson, Erik Tonni, Tomonori Ugajin for collaborations on the
related works. TT is grateful to the organizers of the CERN winter
school 2012 on Supergravity, Strings and Gauge Theory for their
hospitality and the present review article is partially based on his
lectures there. TT is also supported by World Premier International
Research Center Initiative (WPI Initiative) from the Japan Ministry
of Education, Culture, Sports, Science and Technology (MEXT). TT is
partly supported by JSPS Grant-in-Aid for Scientific Research
No.\,20740132 and by JSPS Grant-in-Aid for Creative Scientific
Research No.\,19GS0219.

\clearpage

%\appendix
%\section{List of macros for formatting text, figures and tables}

\end{document}